\documentclass{ws-procs975x65}%
\usepackage{amsmath}
\usepackage{amsfonts}
\usepackage{amssymb}
\usepackage{graphicx}%
\begin{document}

\title{Self-sustained traversable wormholes in modified gravity theories}

\author{Remo Garattini$^1$, Francisco S.N. Lobo$^2$}

\address{$^1$Universit\`{a} degli Studi di Bergamo, Facolt\`{a} di Ingegneria,\\ Viale
Marconi 5, 24044 Dalmine (Bergamo) ITALY.\\INFN - sezione di Milano, Via Celoria 16, Milan, Italy\\
\email{remo.garattini@unibg.it}}

\address{$^2$Centro de Astronomia e Astrofísica da Universidade de Lisboa,\\
Campo Grande, Ed. C8 1749-016 Lisboa, PortugaL\\
\email{flobo@cii.fc.ul.pt
}}

\begin{abstract}
We consider the possibility that wormhole geometries are sustained by their
own quantum fluctuations, in the context of noncommutative geometry
and Gravity's Rainbow models. More specifically, the energy density of the
graviton one-loop contribution to a classical energy in a wormhole background
is considered as a self-consistent source for wormholes. In this
semi-classical context, we consider the effects of a smeared particle-like
source in noncommutative geometry and of the Rainbow's functions in sustaining wormhole
geometries.

\end{abstract}

\bigskip\bodymatter

A traversable wormhole is a solution to the Einstein Field equations,
represented by two asymptotically flat regions joined by a bridge. The spacetime metric
representing a spherically symmetric and static wormhole is given by
\begin{equation}
ds^{2}=-e^{2\Phi(r)}\,dt^{2}+\frac{dr^{2}}{1-b(r)/r}+r^{2}\,(d\theta^{2}%
+\sin^{2}{\theta}\,d\phi^{2})\,,\label{metricwormhole}%
\end{equation}
where $\Phi(r)$ and $b(r)$ are arbitrary functions of the radial coordinate,
$r$, denoted as the redshift function, and the form function, respectively
\cite{MT}. The radial coordinate has a range that increases from a minimum
value at $r_{0}$, corresponding to the wormhole throat, to infinity. A
fundamental property of a wormhole is that a flaring out condition of the
throat, given by $(b-b^{\prime}r)/b^{2}>0$, is imposed \cite{MT,Visser}, and
at the throat $b(r_{0})=r=r_{0}$, the condition $b^{\prime}(r_{0})<1$ is
imposed to have wormhole solutions. In order for the wormhole to be traversable, one must demand
that there are no horizons present, so that $\Phi(r)$ must be finite everywhere. 

Indeed, the flaring out condition, through the Einstein field equation, imposes the violation of the null energy condition. Matter that violates the latter is denoted as {\it exotic matter}. As classical matter satisfies the pointwise energy conditions, it is likely that wormholes must belong to realm of the semiclassical regime, or perhaps to a possible quantum theory of
the gravitational field. Since a viable complete theory of quantum gravity has yet to be formulated, it is important to approach this problem semiclassically. On this ground, the Casimir energy on a fixed background has the correct properties to substitute the exotic matter: indeed, it is known that, for different physical systems, Casimir energy is negative. However, instead of studying the
Casimir energy contribution of some matter or gauge fields in sustaining wormholes, we propose to use the energy of the graviton of a background wormhole geometry. In this way, one can assume that the
quantum fluctuations of the wormholes can be used as a fuel to sustain these exotic spacetimes.

Henceforth, we consider a constant redshift function, $\Phi^{\prime
}(r)=0$, so that the curvature scalar reduces to $R=2b^{\prime}/r^{2}$. Thus, the classical energy is given by
\begin{equation}
H_{\Sigma}^{(0)}=\int_{\Sigma}\,d^{3}x\,\mathcal{H}^{(0)}=-\frac{1}{16\pi
G}\int_{\Sigma}\,d^{3}x\,\sqrt{g}\,R  = -\frac{1}{2G}\int_{r_{0}}^{\infty}\,\frac{dr\,r^{2}}%
{\sqrt{1-b(r)/r}}\,\frac{b^{\prime}(r)}{r^{2}}\,.\label{classical}
\end{equation}
where the background field super-hamiltonian, $\mathcal{H}^{(0)}$, is
integrated on a constant time hypersurface. Now, a traversable wormhole is said to be \textquotedblleft%
\textit{self sustained}\textquotedblright\ if%
\begin{equation}
H_{\Sigma}^{(0)}=-E^{TT},
\end{equation}
where $E^{TT}$ is the total regularized graviton one loop energy, which is given by
\begin{equation}
E^{TT}=-\frac{1}{2}\sum_{\tau}\left[  \sqrt{E_{1}^{2}\left(  \tau\right)
}+\sqrt{E_{2}^{2}\left(  \tau\right)  }\right]  \,.\label{ETT}%
\end{equation}
$\tau$ denotes a complete set of indices and $E_{i}^{2}\left(
\tau\right)  >0$, $i=1,2$ are the eigenvalues of the modified Lichnerowicz
operator%
\begin{equation}
\left(  \hat{\bigtriangleup}_{L\!}^{m}\!{}\;h^{\bot}\right)  _{ij}=\left(
\bigtriangleup_{L\!}\!{}\;h^{\bot}\right)  _{ij}-4R{}_{i}^{k}\!{}h_{kj}^{\bot
}+\text{ }^{3}R{}\!{}h_{ij}^{\bot}\,,\label{Lich}%
\end{equation}
acting on traceless-transverse tensors of the perturbation and where
$\bigtriangleup_{L}$is the Lichnerowicz operator defined by%
\begin{equation}
\left(  \bigtriangleup_{L}\;h\right)  _{ij}=\bigtriangleup h_{ij}%
-2R_{ikjl}\,h^{kl}+R_{ik}\,h_{j}^{k}+R_{jk}\,h_{i}^{k},
\end{equation}
with $\bigtriangleup=-\nabla^{a}\nabla_{a}$. When we perform the sum over all
modes, $E^{TT}$ is usually divergent. In this context, different
models which use phantom energy \cite{Remo,RGFSNL}  and noncommutative geometry \cite{RGFSNL2} 
as an additional source of negative energy have been proposed. However, all of them use a zeta
regularization and a renormalization scheme to handle the divergences. 

In fact, it has been shown that noncommutativity eliminates point-like structures in favor of smeared 
objects in flat spacetime. Thus, one may consider the possibility that noncommutativity could cure the 
divergences that appear in general relativity. The effect of the smearing is mathematically implemented 
with a substitution of the Dirac-delta function by a Gaussian distribution of minimal length. In 
particular, the energy density of a static and spherically symmetric, smeared and particle-like 
gravitational source is given by
\begin{equation}
\rho_{\alpha}(r)=\frac{M}{(4\pi\alpha)^{3/2}}\;\mathrm{exp}\left(
-\frac{r^{2}}{4\alpha}\right)  \,, \label{NCGenergy}%
\end{equation}
where the mass $M$ is diffused throughout a region of linear dimension
$\sqrt{\alpha}$ due to the intrinsic uncertainty encoded in the coordinate commutator. In fact, wormhole geometries were analyzed \cite{RGFSNL2} considering that the equation governing quantum fluctuations behaves as a backreaction equation. In particular, the energy density of the graviton one loop contribution to a classical energy in a wormhole background and the finite one loop energy density is considered as a self-consistent source for these wormhole geometries. Furthermore, interesting solutions were found for an appropriate range of the parameters, validating the perturbative computation introduced in this semi-classical approach. We refer the reader to \cite{RGFSNL2} for more details.

An interesting alternative has been proposed in \cite{RGFSNLMDR} which adopts
Gravity's Rainbow\cite{MS} as a UV regulator. Basically, one introduces two
unknown functions $g_{1}\left(  E/E_{P}\right)  $ and $g_{2}\left(
E/E_{P}\right)  $ which have the following property%
\begin{equation}
\lim_{E/E_{P}\rightarrow0}g_{1}\left(  E/E_{P}\right)  =1\qquad\text{and}%
\qquad\lim_{E/E_{P}\rightarrow0}g_{2}\left(  E/E_{P}\right)  =1\label{lim}%
\end{equation}
and the wormhole line element is distorted to become
\begin{equation}
ds^{2}=-N^{2}\left(  r\right)  \frac{dt^{2}}{g_{1}^{2}\left(  E/E_{P}\right)
}+\frac{dr^{2}}{\left(  1-\frac{b\left(  r\right)  }{r}\right)  g_{2}%
^{2}\left(  E/E_{P}\right)  }+\frac{r^{2}}{g_{2}^{2}\left(  E/E_{P}\right)
}\left(  d\theta^{2}+\sin^{2}\theta d\phi^{2}\right)  \,.\label{dS}%
\end{equation}
This implies that the Lichnerowicz operator $\left(  \ref{Lich}\right)  $
is also distorted and the one loop contribution $\left(  \ref{ETT}\right)  $ can
give finite contribution for appropriate choices of $g_{1}\left(
E/E_{P}\right)  $ and $g_{2}\left(  E/E_{P}\right)  $ avoiding therefore any
regularization/renormalization scheme. In this semiclassical context, specific choices were considered for the rainbow’s functions and solutions for wormhole geometries were found in the cis-Planckian and trans-Planckian regimes \cite{RGFSNLMDR}. In the former regime, the wormhole spacetimes are not asymptotically flat and need to be matched to an exterior vacuum solution \cite{DeBenedictis:2008qm}. In the latter trans-Planckian regime, the quantum corrections are exponentially suppressed, which provide asymptotically flat wormhole geometries. Furthermore, one can even adopt a noncommutative
geometry strategy developed on the phase space such as in \cite{RGPN,RGTM2012} but unfortunately one finds that the wormholes are traversable in principle but not in practice.

\end{document}